\documentclass{kapproc} 
\usepackage{graphicx}
\usepackage{subfigure}
\usepackage{amsmath}

\kluwerbib

\let\lcitebracket(
\let\rcitebracket)

\begin{document}



\articletitle[]{Collective properties of electrons and holes in
coupled quantum dots }


\author{Guido Goldoni, Filippo Troiani, Massimo Rontani, Devis Bellucci, \\
and Elisa Molinari}
\affil{INFM National Research Center on nanoStructures
and bioSystems at Surfaces (S3)\\
Dipartimento di Fisica, Universit\`a degli Studi di Modena e
Reggio Emilia,\\ Via Campi 213/A, 41100 Modena, Italy}

\author{Ulrich Hohenester}
\affil{Institute f\"ur theoretische physik,
Karl-Franzens-Universit\"at Graz,\\ Universit\"asplatz 5, 8010 Graz,
Austria}



\begin{abstract}
We discuss the properties of few electrons and electron-hole pairs
confined in coupled semiconductor quantum dots, with emphasis on
correlation effects and the role of tunneling. We shall discuss,
in particular, exact diagonalization results for biexciton binding
energy, electron-hole localization, magnetic-field induced Wigner
molecules, and spin ordering.
\end{abstract}


\section{Introduction}

The similarity between quantum dots (QDs) and natural atoms,
ensuing from the discrete density of states, is often pointed out
\cite{MaksymPRL,Kastner,Ashoori96,Tarucha96,RontaniNato}.
Shell structure \cite{Ashoori96,Tarucha96}, correlation effects
\cite{BryantPRL}, and Kondo physics \cite{Goldhaber-Gordon98,Cronenwett98}
are among the most striking demonstrations. Aside from these similarities,
and in addition to the huge technological interest in quantum
systems which can be grown with very high control, there exist two
main differences between natural and artificial atoms that make
QDs particularly interesting also from a fundamental point of
view. First, while in natural atoms Coulomb interactions are
typically of the same order of the lowest energy single-particle
gaps, in QDs these interactions can be made larger or much larger
than the latter, due to the different scaling of kinetic and
Coulomb energy contributions with respect to confinement length.
Secondly, also magnetic field daily attainable in laboratories are
associated to energy scales which are comparable with
single-particle gaps; therefore, they can be used to manipulate
the quantum states and, in particular, the ratio between kinetic
and Coulomb contributions, so that in QDs one can reach regimes
which are unattainable in natural atoms.

Coupled quantum dots, also called Artificial Molecules (AMs),
extend to the molecular realm the similarity between quantum dots
and artificial atoms
\cite{LeoLattice,LivermoreScience,Blick98a,Brodsky00,Rontani01,Pi01}.
Inter-dot tunneling introduces a kinetic contribution which has to
be added to the QD confinement energy, and can be tuned with
respect to it by structure engineering and by transverse magnetic
fields. In addition, Coulomb interactions, both inter-dot and
intra-dot, mix up in a very complicated way different electronic
configurations (Slater determinants), the ratio between different
contributions being a function, for example, of the external
magnetic field strength and direction. Therefore, AMs constitute
an interesting laboratory to study the physics of correlation. A
specific aspect of AMs, with respect to single QDs, is that
carrier-carrier interaction is tied with localization. Indeed,
while kinetic energy favors delocalization over e.g.~two QDs
forming a diatomic AM, Coulomb interaction favors localization in
opposite dots for charges of the same sign; the ratio between the
two contributions and, therefore, the localized or delocalized
character of the correlated state, is controlled by the tunneling
energy.

From the theoretical point of view, highly correlated quantum
systems are obviously a challenge. In many istances one is not
allowed to use mean-field methods
\cite{Pfannkuche93,Szafran03prb}. Reliable results can be obtained
by a Configuration Interaction approach, where one uses the
calculated single-particle states to form a large basis of Slater
determinants to represent the Coulomb interaction. Obviously, this
method has limitations in the number of free-carriers which can be
treated and of single-particle levels included in the basis set.
On the other hand, it provides accurate results which can be
quantitatively compared with experiments and which represent a
benchmark for more approximate methods.

In this work we review some theoretical results obtained for few
electrons and electron-hole pairs in AMs. We specifically focus on
the effects of carrier-carrier interactions, and the interplay of
tunneling with other energy scales. All calculations are obtained
by exact diagonalization methods of the Coulomb interaction. In a
typical calculations we consider $N_e$ electrons and $N_h$ holes
with the effective-mass Hamiltonian
\begin{equation}\label{eq:Hamiltonian}
    H  =  H_e + H_h + H_{eh},
\end{equation}
where
\begin{eqnarray}
   H_\alpha & = & \sum_{i=1}^{N_\alpha} \left[-\frac{\hbar^2}{2m^*_\alpha}
    \left({\nabla}_i + \frac{e}{c}\mathbf{A}(\mathbf{r}^{\alpha}_i)\right)^2
     + V_\alpha(\mathbf{r}^{\alpha}_i)\right]\nonumber\\
     & + & \frac{1}{2}\sum^{N_\alpha}_{\substack{i,j=1 \\ i\neq j}} \frac{e^2}{\epsilon^*
    |\mathbf{r}^{\alpha}_i-\mathbf{r}^{\alpha}_j|^2}
     + g^*\mu_B \vec{\sigma}\cdot \mathbf{B} ,\nonumber \\
   H_{eh} & = &  - \sum_{i=1,j=1}^{N_e,N_h}\frac{e^2}{\epsilon^*
    |\mathbf{r}^e_i-\mathbf{r}^h_j|^2},
\label{eq:Hamiltonian2a}
\end{eqnarray}
with $\alpha = e,h$. Here $m^*_\alpha$ and $\epsilon^*$ are the
effective mass and dielectric constant, respectively, $g^*$ is the
effective giromagnetic factor, $\mu_B$ the Bohr magneton,
$\mathbf{A}$ is the vector potential generating the magnetic field
$\mathbf{B}$, and $e=|e|$; all parameters are taken for the AlGaAs
class of materials. Equations (\ref{eq:Hamiltonian2a}) neglect
non-parabolicity and spin-orbit effects, but otherwise describe
samples with realistic properties, such as layer width and finite
band offsets, by means of the effective potentials
$V_\alpha(\textbf{r})$. Typical samples that we shall consider
consist of two identical QDs obtained from symmetric coupled
quantum wells, grown, say, along the $z$ direction, of width
$L_W$, separated by a barrier $d$, and with band-offset $V_0$
between wells and barriers. The lateral confinement can be often
taken as a two-dimensional (2D) parabolic potential characterized
by an energy $\hbar\omega_0$ which may assume rather different
values, depending whether QDs are obtained by gating or by etching
of a 2D electron/hole gas, by self-assembly, etc. The lateral
confinement energy $\hbar\omega_0$ is typically of the order 1-10
meV and often much weaker than the confinement along the growth
direction.

Our numerical approach consists in mapping the single-particle
terms $H_\alpha$ in a real-space grid, leading to a large sparse
matrix which is diagonalized by Lanczos methods. Single-particle
spin-orbitals thus obtained are then used to build a basis of
Slater determinants for the $N$-particle problem, which is then
used to represent the two-body term, in the familiar Configuration
Interaction approach. Coulomb matrix elements are calculated
numerically. The ensuing matrix, which can be very large, is again
sparse and can be diagonalized via the Lanczos method
\cite{ARPACK}.

The paper is organized as follows. In Sec.\ \ref{sec:biexcitons}
we discuss electron-hole complexes, and particularly biexciton
binding and localization. In Sec.\ \ref{sec:electrons} we discuss
the phase diagram of few electrons in a magnetic field. In
particular, in Sec.\ \ref{sec:electrons:wigner} we consider the
effects of tunneling on few electrons in a large vertical field,
where carriers are localized in the so-called Wigner molecule.
Finally, in Sec.\ \ \ref{sec:electrons:inplane} we show how
spin-ordering of two electrons can be manipulated by a magnetic
field of arbitrary strength and direction.

\section{Electron-hole complexes in AM}
\label{sec:biexcitons}

In this section we are concerned with the biexciton states in two
identical vertically-coupled QDs: the main focus is on the effects
of the Coulomb-induced interdot correlations, which are
investigated as a function of the barrier width $d$
\cite{FilippoBiexciton}. The features of
the biexcitonic states are shown to critically depend on the
detailed balance between three phenomena: $(i)$ the interdot
tunneling, which is a single-particle feature, tending to
delocalize the carriers and to spread them over the artificial
molecule; $(ii)$ the homopolar Coulomb interactions, which is a
few-particle effect, resulting in spatial correlations between the
two electrons (holes) that tend to minimize the repulsion energy
by localizing the identical carriers in opposite dots; $(iii)$ the
heteropolar Coulomb interactions, that induce spatial correlations
between electrons and holes in order to maximize their overlap.
Interestingly enough, the interplay between these trends can be
widely tuned, either by modifying the structural parameters of the
dots or by applying external fields, in order to induce
non-trivial behaviours already at the few-particle level.

\begin{figure}
  \centering
  \begin{minipage}[b]{0.45\textwidth}
     \centering
     \includegraphics[width=7truecm]{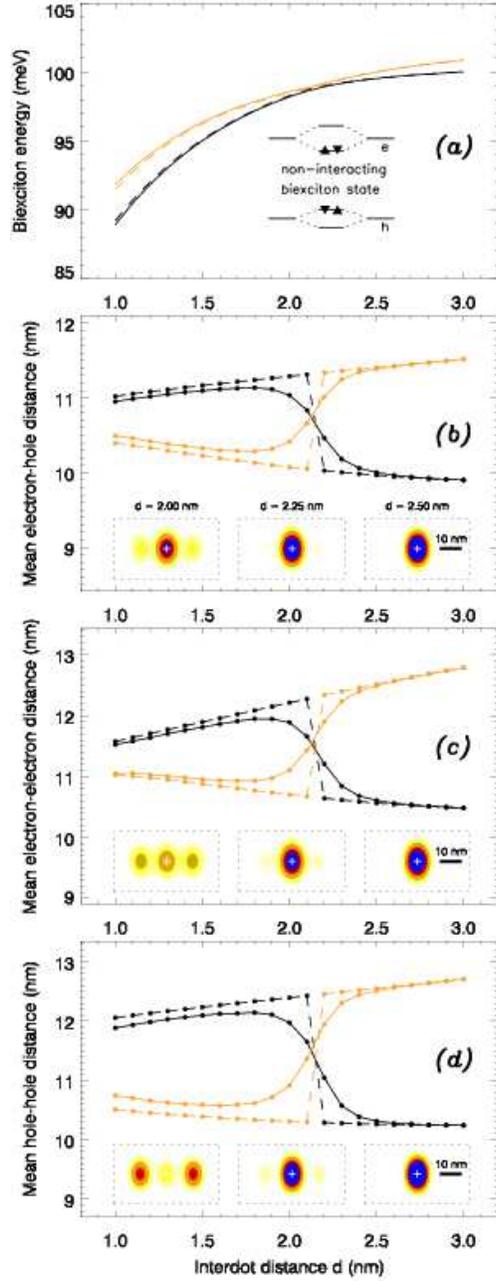}
  \end{minipage}\hfill
  \begin{minipage}[b]{0.45\textwidth}
     \centering
     \caption{Biexciton energies (a) and mean distances between carriers
(b-d) as a function of the interdot distance $d$. The plots refer
to the four states lowest in energy, with even (continuous lines)
or odd (dotted lines) parity. The insets (b-d) show the dependence
of the pair-correlation functions on the relative position ${\bf
r} = {\bf r}_1 - {\bf r}_2$ of the two carriers: in particular, we
take $y=0$ and show the dependence on the $z$ (horizontal axis)
and $x$ (vertical axis) coordinates. The two vertically-coupled
GaAs/AlGaAs QDs are modeled by means of a confinement potential
which is double-well like in the growth direction (the well width
and band offsets are $L_W = 10$~nm and $V_0^{(e,h)}=400,215$~meV
respectively) and parabolic in the plane
($\hbar\omega_0^{(e,h)}=20, 3.5$~meV, $m^*_{(e,h)}=0.067,0.38\;
m_0$).}
     \label{fig:fig-filippo}
  \end{minipage}
\end{figure}

In the following we fix all the physical parameters of the AM but
the interdot distance $d$ in order to explore the different
coupling (tunneling) regimes. At the smallest interdot distance,
the symmetric-antisymmetric (S-AS) splitting
$2t^{(e,h)}=\epsilon_{AS}^{(e,h)}-\epsilon_S^{(e,h)}$ is
maximized, where $\epsilon_S$ ($\epsilon_{AS}$) is the energy of
the single-particle S (AS) level. For the sample of Fig.\
\ref{fig:fig-filippo}, for example, where $d$ is varied in the
range $1\div 3$ nm, the tunneling splitting amounts to about 12
and 6 meV for the electrons and the holes, respectively, at $d =
1$ nm. As a result, both electrons are frozen in the S state and
the spatial distribution of each electron is uncorrelated both
with respect to the other electron position and to that of the two
holes. Due to their smaller effective mass, the holes tunnel less
efficiently than the electrons ($t^{(h)}<t^{(e)})$: therefore the
interdot spatial correlation, which requires the occupation of the
AS states, is favored and acts in such a way that the carriers are
always localized in different dots. This is shown in the left-hand
insets of Fig.\ \ref{fig:fig-filippo}(b-d), where we plot the
dependence of the pair-correlation on the relative position
between the carriers. By progressively increasing the barrier
width, the effects of the homopolar Coulomb interactions between
the carriers continuously increase, whereas electrons and holes
are not correlated with each other. In this kind of configuration,
resulting in a factorized wavefunction $\psi ({\bf r}_{e1},{\bf
r}_{e2},{\bf r}_{h1},{\bf r}_{h2}) \simeq \phi^{(e)} ({\bf
r}_{e1},{\bf r}_{e2}) \phi^{(h)} ({\bf r}_{h1},{\bf r}_{h2})$, the
probability of having double occupancies (two electrons or two
holes) in each dot is strongly suppressed, i.e., the two excitons
are localized in different dots. In approaching the weak-coupling
regime, the biexciton groundstate undergoes a rapid transition
towards a maximally correlated configuration, where all the
carriers are localized in the same dot (see the right-hand insets
of Fig.\ \ref{fig:fig-filippo}(b-d)). The reason why this
arrangement is energetically favored as compared to the previously
discussed one is entirely related to the in-plane correlations
between the carriers. In fact, due to the substantial symmetry of
the electron and hole wavefunctions, the Coulomb energy of the two
configurations is the same for both types of particles in the
mean-field limit: the increase in the Coulomb repulsion arising
from the stronger degree of localization is cancelled by that of
the Coulomb attraction. The occupation of the higher Fock-Darwin
states, however, gives rise to additional (in-plane) spatial
correlations, such as the ones that cause the biexciton binding
energies in single QDs. This kind of interaction, whose closest
classical analogue is an induced dipole-induced dipole force, is a
short range one and is therefore uneffective as far as the two
excitons are localized in different dots. In Fig.\
\ref{fig:fig-filippo} we plot the energy levels [panel (a)] and
the average distance between each pair of carriers [panels
(b)-(d)] for the four lowest biexciton states: the continuous
(dotted) lines correspond to the two states of even (odd)
symmetry. Two features emerge from the plots: $(i)$ the transition
between the weakly- and the highly-correlated configurations
occurs quite abruptly for $d\simeq 2$~nm; $(ii)$ in a limited
interval around this value each of the four states swaps its
features with state of equal symmetry, and correspondingly an
anticrossing occurs between their energy levels.

The interdot correlation is therefore seen to play a crucial role
in determining the carrier localization in artificial molecules:
these exhibit a fully three-dimensional nature, where novel
behaviors occur as compared to the single dots. Indeed, such
features have to be taken into account within the design of
exciton-based quantum computation schemes and devices
\cite{FilippoQIP}, where the
double occupancies are known to spoil the required (tensorial)
Hilbert-space structure.

\section{Interacting electrons in AM in a magnetic fields}
\label{sec:electrons}

\begin{figure}
  \subfigure[]{
      \begin{minipage}[b]{0.5\textwidth}
         \includegraphics[width=5truecm]{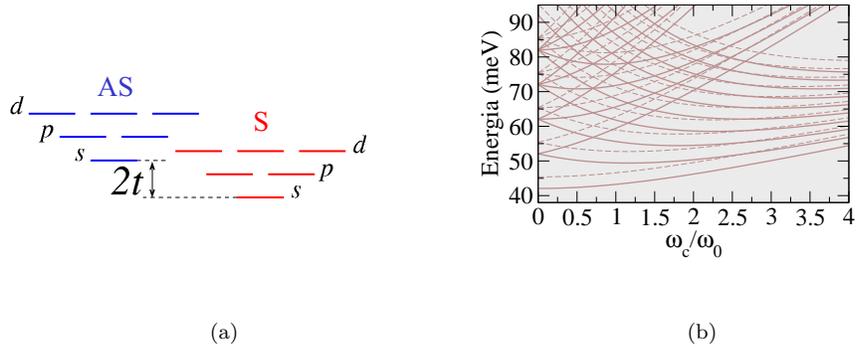} \\
         \vspace{5truemm}
      \end{minipage}
      \label{fig:single_particle:a}}
  \subfigure[]{
      \begin{minipage}[b]{0.5\textwidth}
         \includegraphics[width=5truecm]{fd_states.eps} \\
         \vspace{-2truemm}
      \end{minipage}
      \label{fig:single_particle:b}}
  \caption{(a) Sketch of the energy levels in a 2D parabolic potential
  for an AM at zero magnetic field.
  $s, p, d$ label the value of the single-particle magnetic moment
  consistent with the corresponding atomic notation. S and AS
  stand for Symmetric and AntiSymmetric levels arising from two
  coupled QDs with tunneling energy $2t$.
  (b) Fock-Darwin levels (Eq.\ \ref{eq:single_particle})
  for an AM with $L_w = 10$ nm, $d = 3$ nm, $V_0 = 300 $ meV, and
  $\hbar\omega_0= 10$ meV.}
  \label{fig:single_particle}
\end{figure}

In this section we shall consider the effects of a magnetic field
of arbitrary strength and direction on the few-electron states in
AMs. Let us first summarize a few well-known effects of a field
which is parallel to the axis of the AM.

Figure \ref{fig:single_particle:a} shows the single-particle
levels at zero field, separated by $\hbar\omega_0$, ensuing from
the 2D parabolic potential of the in-plane confinement. The two
sets of shells are separated by the tunneling energy. As a
vertical field is switched on, the Hamiltonian can still be
analytically solved \cite{Jacak}, and gives rise to the Fock-Darwin (FD) levels
\begin{equation}
\varepsilon^i_{nm} = \varepsilon_i + \hbar\Omega
(2n+|m|+1) -(\hbar\omega_c/2) m. \label{eq:single_particle}
\end{equation}
Here $n$ is the principal quantum number describing the radial
distribution of the wavefunction, and $m$ the azimuthal quantum
number labelling the angular momentum which is conserved in this
cylindrically symmetric configuration. The oscillator frequency
$\Omega = (\omega_0^2+\omega_c^2/4)^{1/2}$, with the cyclotron
frequency $\omega_c= eB/m^{*} c$, shows the competition between
the confinement energy gaps, $\hbar\omega_0$, and the
field-induced gaps, $\hbar\omega_c$. The energy $\varepsilon_i$ is
the confinement energy along the growth direction. For AMs with
sufficiently narrow symmetric quantum well we can limit our
considerations to two levels, which are the ground S and AS states
($i$ = S, AS). The Fock-Darwin states for an AM are shown with
lines in Fig.\ \ref{fig:single_particle:b}. At zero field one
recovers the $s,p,...$ shells of Fig.\
\ref{fig:single_particle:a}. At finite field, the $\pm m$ orbital
degeneracy splits, and at large fields the levels with highest $m$
tend to form highly degenerate Landau levels.

\begin{figure}
  \centering
  \includegraphics[width=7truecm]{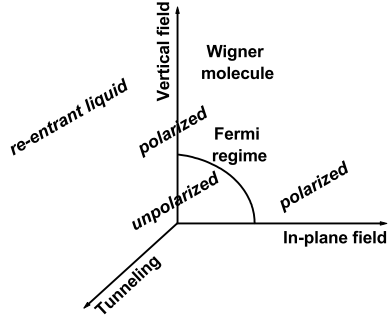}\\
  \caption{Sketch of the ground state phase diagram for an AM with respect
  to field strength and direction, and tunneling.}\label{fig:full_phase_diagram}
\end{figure}

Figure \ref{fig:full_phase_diagram} shows a qualitative sketch of
the ground state phase diagram of few-electrons in a AM which will
be discussed in the next sections. Let us focus, for the time
being, on the vertical field effects. When a field is applied
parallel to the growth axis, the effect is to squeeze the states
in the QD plane. By doing so, Coulomb energy increases and, at
sufficiently large field, the system becomes spin-polarized in
order to gain in exchange energy. At even larger fields, kinetic
energy is quenched, since all single-particle levels tend to
become degenerate [Fig.\ \ref{fig:single_particle:b}], and
correlation dominates the system. Accordingly, the electron system
first becomes a spin-polarized confined Fermi sea, laterally
delocalized over the QDs. Then, at very high fields, the systems
transforms to a confined Wigner crystal, in which the carriers are
localized in the plane of the QD; the latter regime is often
called Wigner molecule \cite{MaksymReview}.

There are two obvious directions along which the phase diagram of
a AM can be extended: the tunneling energy and an in-plane
component of the magnetic field, as shown in Fig.\
\ref{fig:full_phase_diagram}. These two regions will be discussed
next.

\subsection{Wigner molecules and tunneling}
\label{sec:electrons:wigner}

\begin{figure}
  \centering
  \includegraphics[width=8truecm]{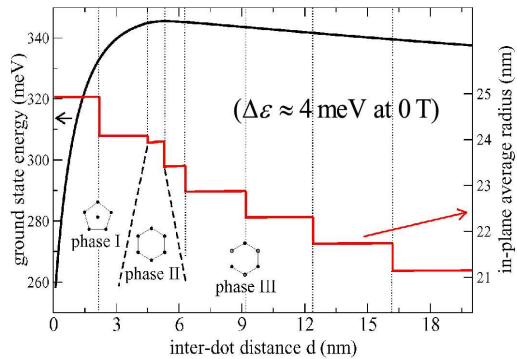}\\
  \caption{Ground-state energy (left axis) and
in-plane average radius $\left<\varrho\right>$ (right axis) vs
inter-dot distance $d$ for six electrons at $B=25\,\mbox{T}$.
Sample parameters are: $\hbar\omega_0 = 3.7$ meV, $V_0 = 250$ meV,
and $L_W = 12$ nm. Insets show the electron arrangements in the
different phases.}\label{fig:wigner_molecule}
\end{figure}

\begin{figure}[h]
\setlength{\unitlength}{1mm}
\begin{picture}(120,125)
\put(0,0){\includegraphics[angle=0,width=8.5cm]{angular-NATO.eps}}
\put(93,92){\includegraphics[angle=0,width=2.5cm]{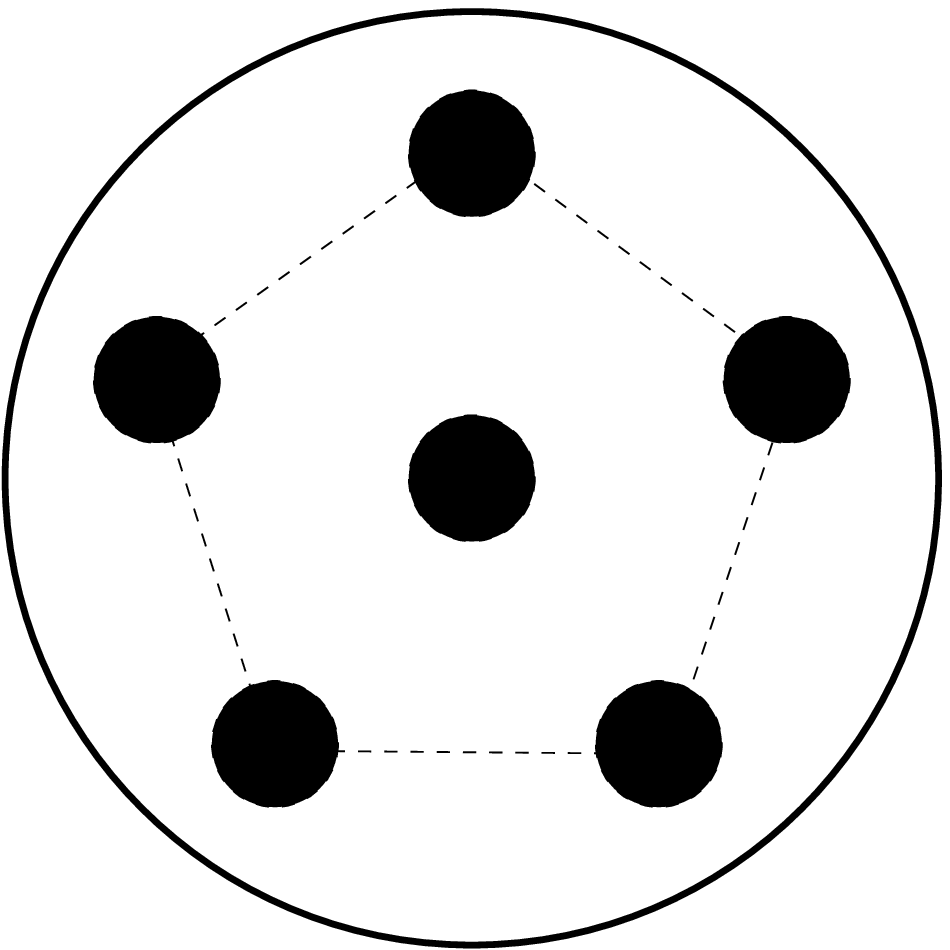}}
\put(95,87){\LARGE $\Delta\varepsilon\ll t$}
\put(93,55){\includegraphics[angle=0,width=2.5cm]{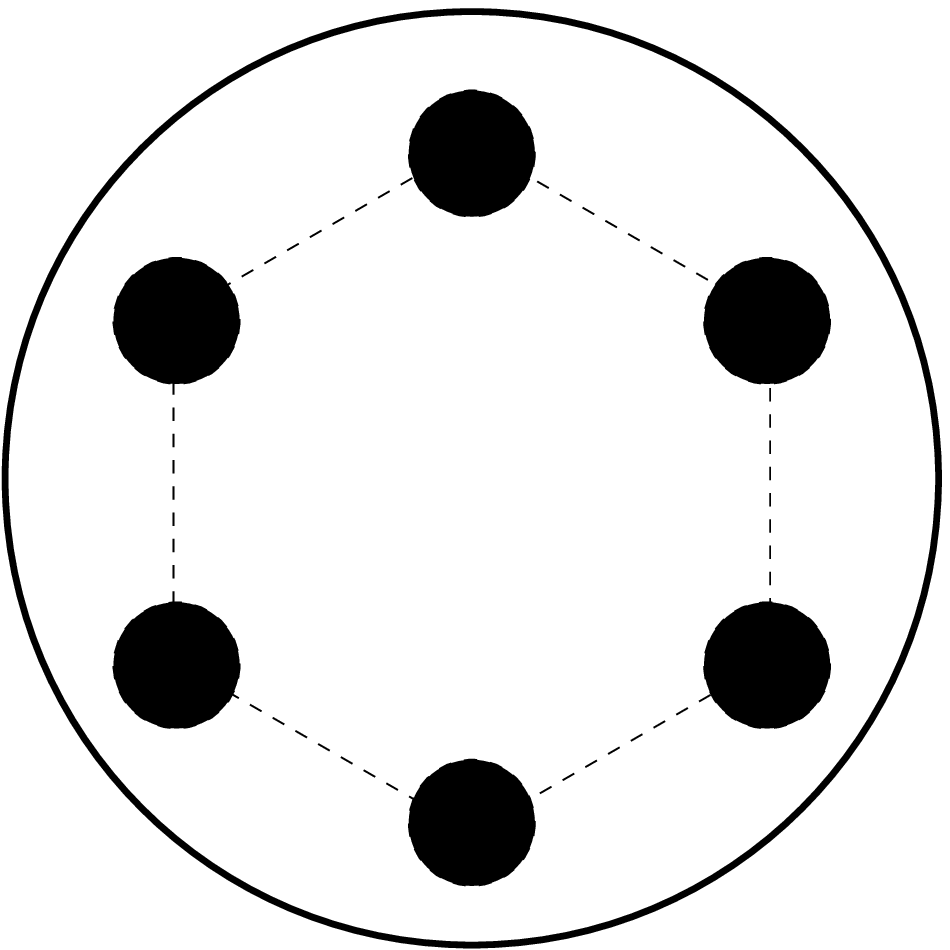}}
\put(95,50){\LARGE $\Delta\varepsilon\approx t$}
\put(93,18){\includegraphics[angle=0,width=2.5cm]{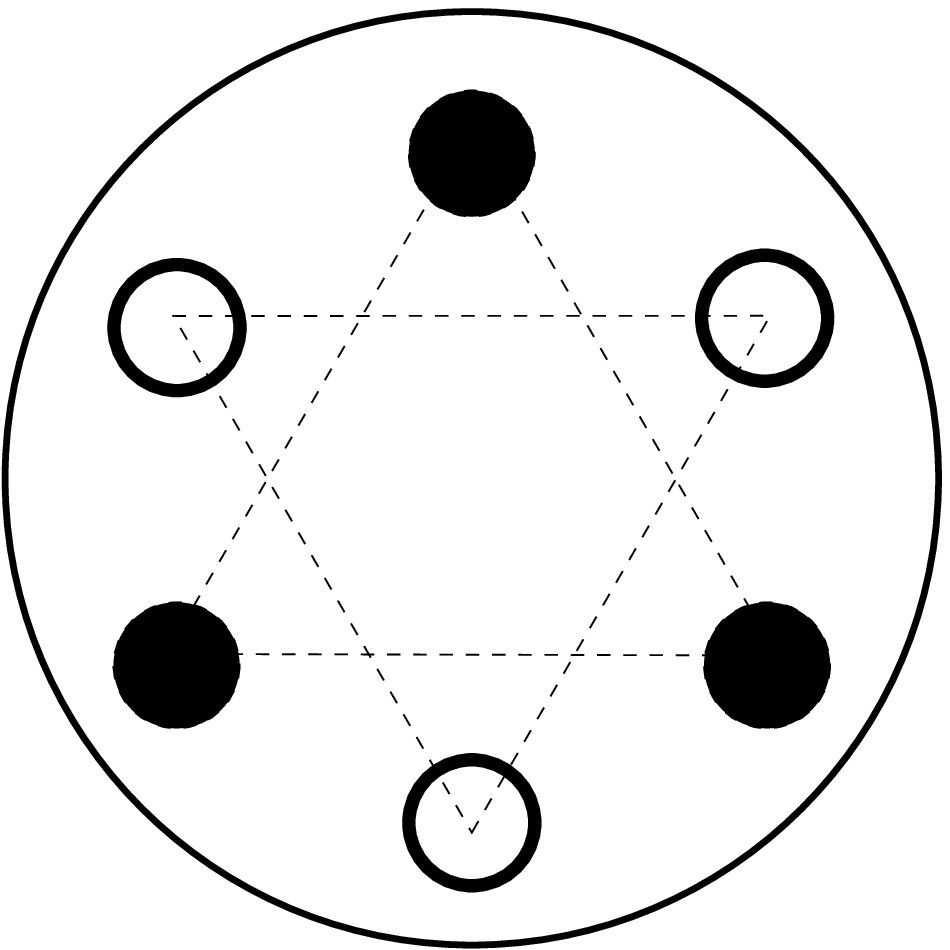}}
\put(95,13){\LARGE $\Delta\varepsilon\gg t$}
\put(50,89){\includegraphics[angle=0,width=4.0cm]{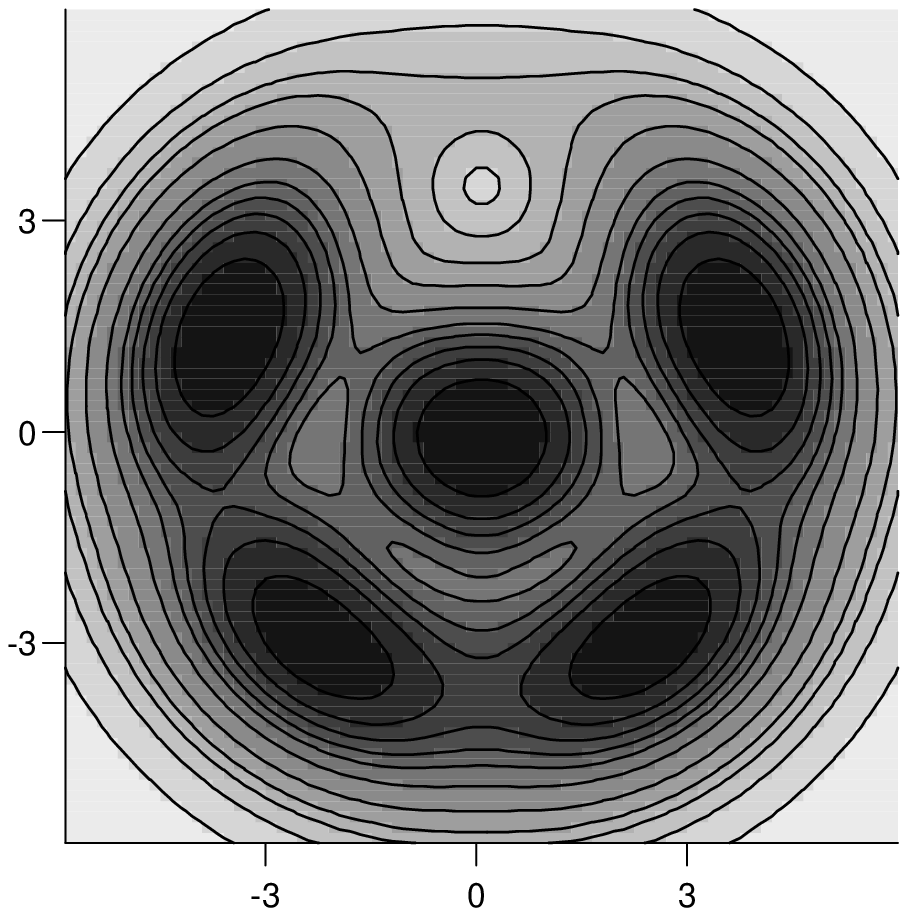}}
\put(77,91){$x$} \put(53,115){$y$}
\put(59.5,96.5){\includegraphics[angle=0,width=1.6cm]{tondo.eps}}
\end{picture}
\caption{Angular correlation function of the three phases of Fig.\
\ref{fig:wigner_molecule}. The three phases correspond to interdot
distances $d$=2, 4.6, 8 nm, respectively.
}\label{fig:angular_correlation}
\end{figure}

As an example of the effects of tunneling in the high field
regime, we show in Fig.\ \ref{fig:wigner_molecule} the calculated
ground-state energy vs the inter-dot distance $d$ for six
electrons \cite{Rontani02}. According to the previous discussion,
we assume that at this field carriers are spin polarized;
furthermore, at the high field considered here, electrons in a
single QD are expected to be localized, and quantum fluctuation to
play a minor effect. In the inset we also show the geometrical
configuration of the localized carriers which will be discussed in
more detail in Fig.\ \ref{fig:angular_correlation}. At small
distances the energy increases with $d$ (phase I), because the
kinetic energy exponentially grows due to the progressive
localization of the wavefunction into the dots: electrons occupy
only S orbitals, whose energies increase. At these small $d$ the
AM behaves as a single QD, and electrons sit at the vertices of a
centered pentagon, which minimizes the Coulomb energy, as
predicted by a classical static calculation \cite{Partoens97}.
There is another obvious regime: when the barrier is large, the
six electrons sit, three per QD, at the vertices of two triangles
staggered by $60^\circ$ (Phase III). Close to $d=5$ nm  inter-dot
tunneling stabilizes Phase II, in which electrons sit at the
vertices of an hexagon. The figure also shows the calculated
average radius $\left<\varrho\right>$ [$\mbox{\boldmath
$\varrho$}\equiv(x,y)$]; Fig.~\ref{fig:wigner_molecule} shows that
these states are incompressible, in the same sense as Laughlin's
states of the FQHE \cite{bob}. Indeed, varying $d$ acts like an
external pressure applied in the $z$ direction, forcing the
wavefunction to change: however, due to a cusp-like structure of
the energy spectrum \cite{MaksymPRL}, this happens only in a
discontinuous way, except for Phase II. We show below that Phase
II has very special properties and is stabilized by tunneling
fluctuations.

To analyze the ground state of the artificial molecule in the
different regimes, we show in Fig.~\ref{fig:angular_correlation}
the pair correlation function $P(\mbox{\boldmath $\varrho$} , z ;
\mbox{\boldmath $\varrho$}_0 , z_0 ) =\sum_{i\neq j}\left<\delta
(\mbox{\boldmath $\varrho$}- \mbox{\boldmath $\varrho$}_i) \delta
( z - z_i ) \delta (\mbox{\boldmath $\varrho$}_0- \mbox{\boldmath
$\varrho$}_j) \delta (z_0 - z_j) \right>/N_e(N_e-1)$ (the average is
on the ground state). Figure
\ref{fig:angular_correlation} shows $P( \mbox{\boldmath $\varrho$}
, z ; \mbox{\boldmath $\varrho$}_0 , z_0 )$ along a circle in the
same dot (solid line) or in the opposite dot (dashed line) with
respect to the position of a reference electron, taken at the
maximum of its charge density, $(\mbox{\boldmath $\varrho$}_0 ,
z_0)$. The right column shows the electron arrangement in the QDs
as inferred by the maxima of $P( \mbox{\boldmath $\varrho$} , z ;
\mbox{\boldmath $\varrho$}_0 , z_0 )$.

At small $d$ (Phase I) the whole system is coherent, i.e., it
behaves as a unique QD; indeed, the correlation function peaks,
forming the outer shell of the centered pentagon, have the same
height in both QDs. At intermediate values of the tunneling
energy (Phase II) the peaks, corresponding to the vertices of a
regular hexagon, have different heights. Finally, when $d$ is
sufficiently large (Phase III), the structure evolves into two
isolated dots coupled only via Coulomb interaction; accordingly,
the peaks are again of the same height, but shifted by 60$^\circ$ .

It is important to note from Fig.~\ref{fig:angular_correlation}
that Phase I and III are strongly localized phases, as
demonstrated by the high peak-to-valley ratio of the correlation
function, and quantum fluctuations play a minor role; therefore,
electron configurations are basically determined by Coulomb
interactions, and have completely classical counterparts
\cite{Partoens97}. On the contrary, in Phase II tunneling
fluctuations prevent electrons from localizing and the
configuration has a ``liquid'' character. Such phase cannot be
explained in term of Coulomb interactions solely and, in fact, the
hexagonal arrangement shown in Fig.~\ref{fig:angular_correlation}
is classically unstable. Therefore, tunneling fluctuations may
induce melting of the otherwise well localized Wigner molecule in
the high field regime, and induce reentrant liquid phases, as
schematically indicated in Fig.\ \ref{fig:full_phase_diagram}. A
discussion of the possible experimental signatures of the
different phases in inelastic light scattering experiments can be
found in \cite{Rontani02}.

\subsection{Effects of an in-plane magnetic field}
\label{sec:electrons:inplane}

\begin{figure}
  \begin{minipage}[c]{0.4\textwidth}
     \centering\includegraphics[width=4truecm]{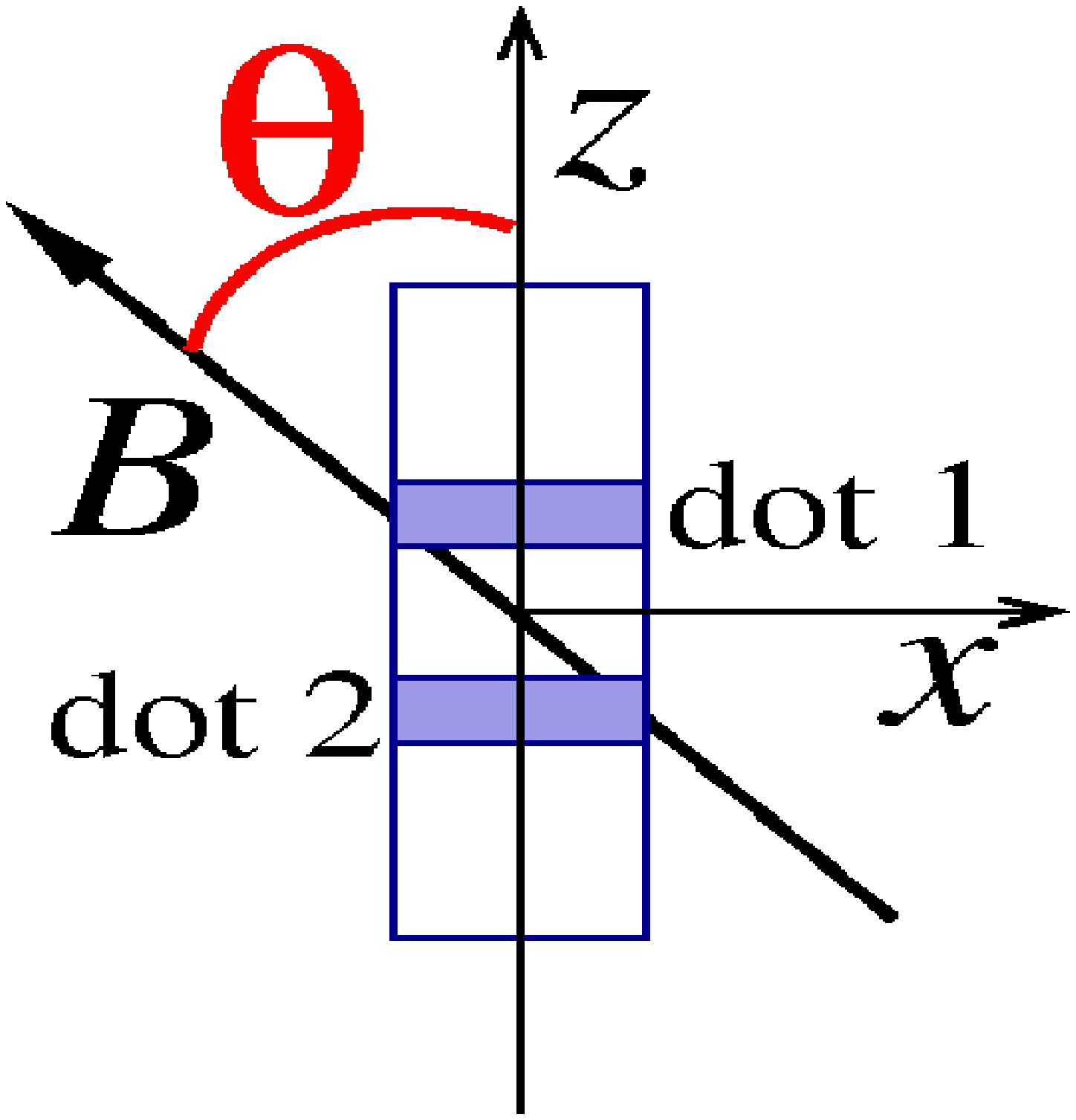}
  \end{minipage}
  \begin{minipage}[c]{0.4\textwidth}
     \centering\includegraphics[width=7truecm]{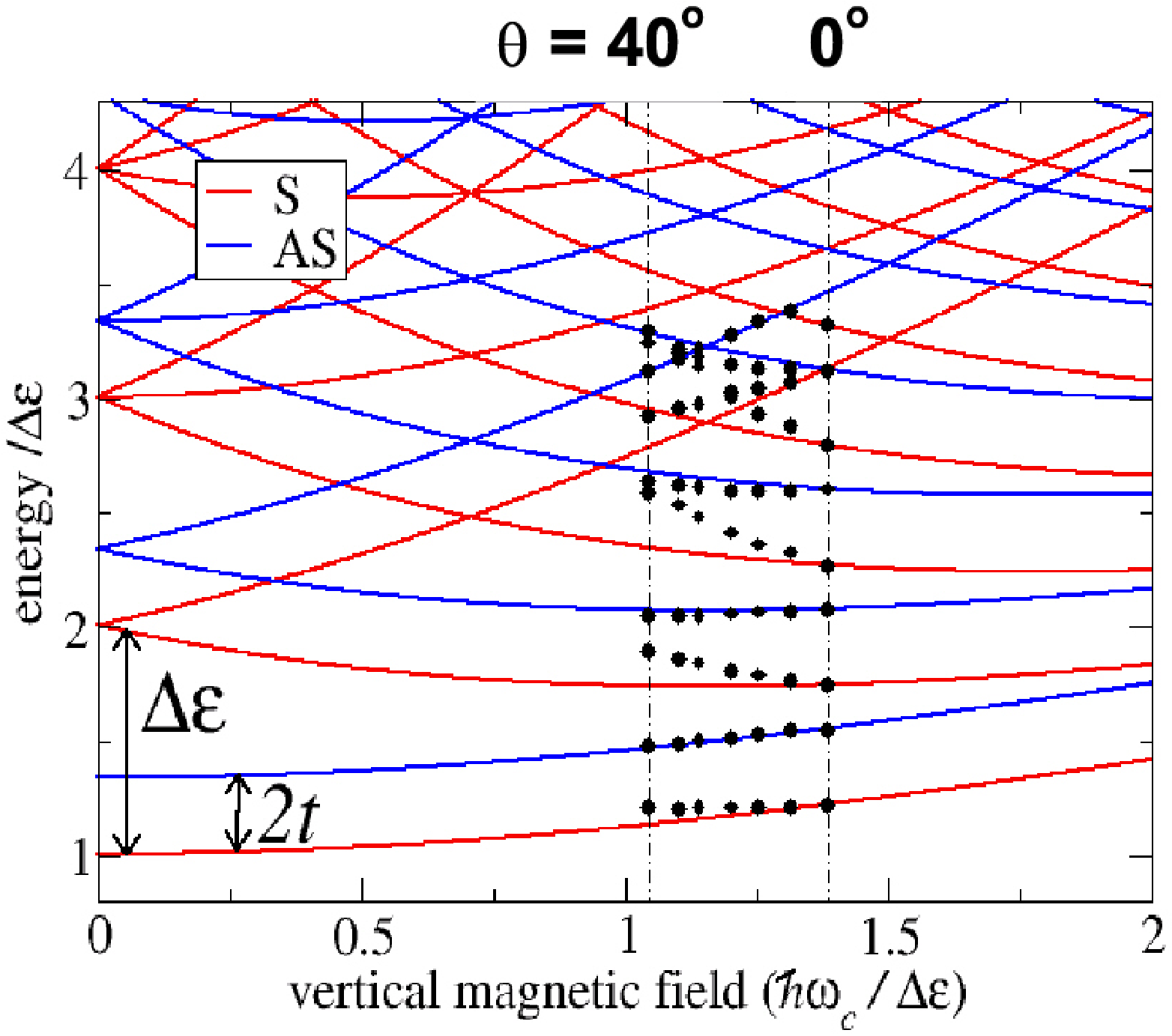}
  \end{minipage}
  \caption{Evolution of single particle states in a AM when the field is
  rotated from the vertical direction by an angle $\theta$. Sample
  parameters are: $L_{w} = 10$ nm, $d = 3$ nm, $V_0 = 300$ meV,
  $\hbar\omega_{0}= 10$ meV. Solid
  lines represent the Fock-Darwin states
  $\varepsilon^i_{nm}$ induced by a strictly vertical field.
  \label{fig:rotated_field}}
\end{figure}

Next we discuss the effect of a finite in-plane component of the
field, $B_\parallel$. To show the effect on the single-particle
states, let us suppose that the total field is tilted by an angle
$\theta$ with respect to the growth direction, so that
$B_\parallel \neq 0$. As shown in Fig.~\ref{fig:rotated_field},
with increasing $\theta$ energy levels tend to follow the FD
states backward, which are shown for comparison, since of course
$B_\perp$ decreases with increasing angle; in addition, however,
the splitting between S and AS levels decreases. This shows that
an in-plane component of the field suppresses tunneling. Note that
this effect is larger for higher levels. Note also that here the
S/AS labelling is used for brevity; obviously, for a general field
direction with respect to the tunneling direction wavefunctions do
not have a well defined S/AS symmetry. It is important to stress
that deviations from the FD states are expected because the energy
scale associated with the in-plane field is comparable to the
tunneling gap; for single QDs, for example, reasonable in-plane
field will not affect the FD states, since the in-plane field
energy scale is typically much smaller than the single-particle
gaps induced by the quantum well confinement.

\begin{figure}
  \centering
  \subfigure[]{
      \begin{minipage}[b]{0.3\textwidth}
         \ \hspace{-12truemm}\includegraphics[width=50truemm]{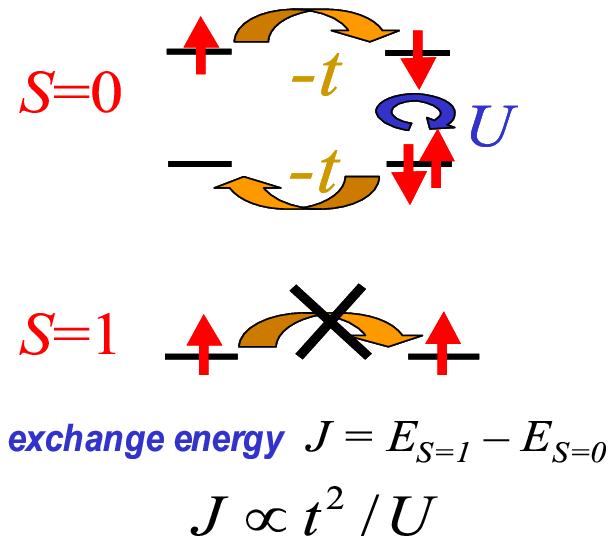}
      \end{minipage}
      \label{fig:exchange:a}}\hfill
  \subfigure[]{
      \begin{minipage}[b]{0.6\textwidth}
         \includegraphics[width=90truemm]{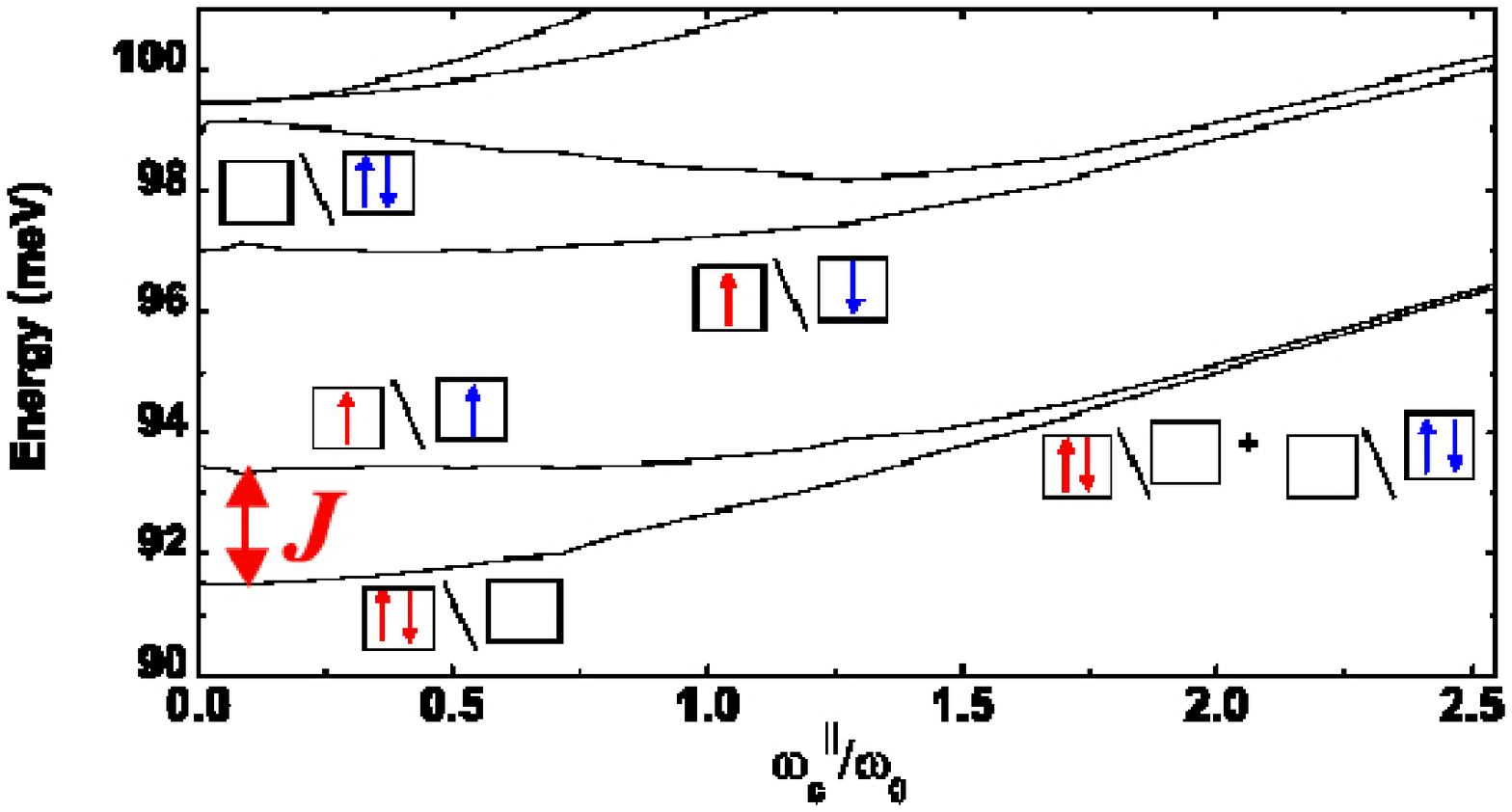}\\
         \vspace{-5truemm}
      \end{minipage}
      \label{fig:exchange:b}}
  \caption{(a) Sketch of the energy contributions to the
   tunneling-induced spin-spin interaction. (b) Two-electron levels,
   with indication of the main component of the wavefunctions in
   terms of S (left boxes) and AS (right boxes) states. Parameters are:
   $L_{w} = 10$ nm, $d = 3$ nm, $V_{0} = 300$ meV, and $\hbar\omega_{0} = 10$
   meV. }\label{fig:exchange}
\end{figure}

In AMs carriers sitting on either dot are not only
electrostatically coupled, but also have their spin interlaced
when tunneling is allowed \cite{Burkard00}. This is sketched in
Fig. \ref{fig:exchange:a}. For two electrons in a singlet state it
is possible to tunnel into the same dot. By doing so, they gain
the tunneling energy $t$; this may compensate for the loss in the
Coulomb energy $U$. This process is forbidden for two electrons in
the triplet state by Pauli blocking. Different spin orderings,
therefore, are associated to an exchange energy, $J$, which to lowest
order in pertubation theory is $J\propto t^2/U$. While in real
molecules $J$ is fixed by the bond length, in AMs it is possible
to tune almost all energy scales by sample engineering and external
fields. Since an in-plane field affects tunneling, we expect the
exchange energy to be affected by an in-plane field as well.

Guided by the above considerations, we next consider the
two-electron system \cite{Bellucci03prl,Bellucci03physe}.
At low vertical fields the ground state of
single and coupled QDs is known to be a singlet state
\cite{Merkt91,Oh96}. In the moderate field regime, therefore, the
lowest energy levels are nearly unaffected by the rotation except
for the shift due to the reduction of the tunneling energy, with
the singlet state being the lowest.

\begin{figure}
  \centering
  \includegraphics[width=76truemm]{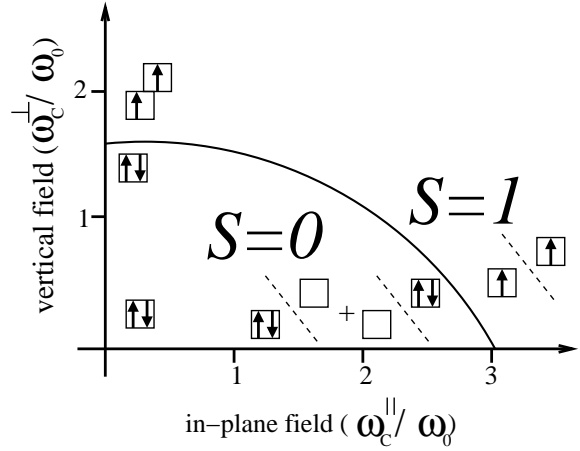}\\
  \caption{Singlet-triplet phase diagram calculated for a GaAs AM
  with $L_W = 10$ nm, $d = 3$ nm, $\hbar\omega_0 =4$ meV.
  The insets show the single-particle occupation in terms of
  S (left to the dashed lines) and AS orbitals (right to the dashed
  lines). At low $B_\parallel$ only S orbitals are occupied.}
  \label{fig:phase_diagram}
\end{figure}

At sufficiently high vertical field a singlet-triplet transitions
take place at a given threshold field. Since the exchange energy
is proportional to tunneling, we expect that the threshold fields
will be lowered as $B_\parallel$ increases \cite{Bellucci03physe}.
This is shown in Fig. \ref{fig:phase_diagram}. The singlet state
is stable in the low field regime. The triplet state becomes
favored in the large field regime, whether the field is
perpendicular or parallel to the plane of the QD; it is should be
noted, however, that this happens by different mechanisms whether
$B_\perp$ or $B_\parallel$ is large. In the former case, the
squeezing of the wavefunction has a Coulomb energy cost which can
only be avoided by triplet spin order. This is analogous to single
QDs. However, while a finite $B_\parallel$ would not affect very
much electronic states in single QDs, where single-particle gaps
are large, in AMs the in-plane field affects the S/AS gap. When
this vanishes, no tunneling energy is lost by localizing in each
dot; then the spin ordering becomes irrelevant, and the triplet
state is favored only due to Zeeman energy
[Fig.~\ref{fig:exchange:b}].

In Figs.\ \ref{fig:exchange:b},\ref{fig:phase_diagram}
we also show in the insets the
character of the two-electron wavefunction of the ground state. In
the low $B_\parallel$ regime, the two electrons occupy only the S
state, either with the $s$ symmetry with opposite spin (low field)
or the $s$ and $p$ symmetry levels with the same spin orientation
(high field), since a large vertical field reduces the $s-p$ gaps.
In the large $B_\parallel$ regime, on the contrary, S and AS
states become degenerate, and are equally occupied by the two
electrons, due to Coulomb correlations.

\begin{figure}
  \centering
  \includegraphics[width=76truemm]{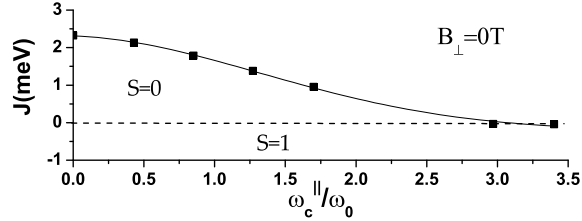}\\
  \caption{Exchange energy $J = E_{S=1}-E_{S=0}$ for a a GaAs AM.
 Same parameters as in Fig. \ref{fig:phase_diagram}.
}
  \label{fig:exchange_energy}
\end{figure}

In Fig. \ref{fig:exchange_energy} we show the exchange energy, $J$,
defined as the difference between the energy of the lowest triplet
and the singlet levels, as a function of the in-plane field at
zero vertical field. This is positive (i.e., the singlet is the
ground state) at low fields, but rapidly decreases as the field
increases. At large fields, the exchange energy changes sign,
being eventually dominated by Zeeman energy.

\section{Acknowledgements}

This work was supported in part by MIUR-FIRB \textsl{Quantum
phases of ultra-low electron density semiconductor
heterostructures}, and by INFM I.T. \textsl{Calcolo Parallelo}
(2003). Con il contributo del Ministero degli Affari Esteri,
Direzione Generale per la Promozione e la Cooperazione
Culturale.






%


\bibliographystyle{apalike}
\chapbblname{procmax}
\chapbibliography{referenze,ReferenzeQD}

\begin{thebibliography}{}

\bibitem[Ashoori, 1996]{Ashoori96}
Ashoori, R. (1996).
\newblock {\em Nature (London)}, 379:413.

\bibitem[Bellucci et~al., ress]{Bellucci03physe}
Bellucci, D., Rontani, M., Goldoni, G., Troiani, F., and Molinari, E. (in
  press).
\newblock Spin-spin interaction in artificial molecules with in-plane magnetic
  field.
\newblock {\em Physica E}.

\bibitem[Blick et~al., 1998]{Blick98a}
Blick, R.~H., Pfannkuche, D., Haug, R.~J., Klitzing, K.~v., and Eberl, K.
  (1998).
\newblock {\em Phys.\ Rev.\ Lett.}, 80:4032.

\bibitem[Brodsky et~al., 2000]{Brodsky00}
Brodsky, M., Zhitenev, N.~B., Ashoori, R.~C., Pfeiffer, L.~N., and West, K.~W.
  (2000).
\newblock {\em Phys.\ Rev.\ Lett.}, 85:2356.

\bibitem[Bryant, 1987]{BryantPRL}
Bryant, G.~W. (1987).
\newblock {\em Phys.\ Rev.\ Lett.}, 59:1140.

\bibitem[Burkard et~al., 2000]{Burkard00}
Burkard, G., Seelig, G., and Loss, D. (2000).
\newblock Spin interactions and switching in vertically tunnel-coupled quantum
  dots.
\newblock {\em Phys.\ Rev.\ B}, 62:2581.

\bibitem[Cronenwett et~al., 1998]{Cronenwett98}
Cronenwett, S.~M., Oosterkamp, T.~H., and Kouwenhoven, L.~P. (1998).
\newblock {\em Science}, 281:540.

\bibitem[Goldhaber-Gordon et~al., 1998]{Goldhaber-Gordon98}
Goldhaber-Gordon, D.~G., Shtrikman, H., Mahalu, D., Abusch-Magder, D., Meirav,
  U., and Kastner, M.~A. (1998).
\newblock {\em Nature (London)}, 391:156.

\bibitem[Jacak et~al., 1998]{Jacak}
Jacak, L., Hawrylak, P., and W\'ojs, A. (1998).
\newblock {\em Quantum dots}.
\newblock Springer, Berlin.

\bibitem[Kastner, 1993]{Kastner}
Kastner, M.~A. (1993).
\newblock {\em Phys.~Today}, 46:24.

\bibitem[Kouwenhoven et~al., 1990]{LeoLattice}
Kouwenhoven, L.~P., Hekking, F. W.~J., van Wees, B.~J., Harmans, C. J. P.~M.,
  Timmering, C.~E., and Foxon, C.~T. (1990).
\newblock {\em Phys.\ Rev.\ Lett.}, 65:361.

\bibitem[Laughlin, 1983]{bob}
Laughlin, R.~B. (1983).
\newblock {\em Phys.\ Rev.\ B}, 27:3383.

\bibitem[Lehoucq et~al., ]{ARPACK}
Lehoucq, R.~B., Maschhoff, K., Sorensen, D.~C., and Yang, C.
\newblock {\em ARPACK computer code}.
\newblock Available at {\tt http://www.caam.rice.edu/software/ARPACK/}.

\bibitem[Livermore et~al., 1996]{LivermoreScience}
Livermore, C., Crouch, C.~H., Westervelt, R.~M., Campman, K.~L., and Gossard,
  A.~C. (1996).
\newblock {\em Science}, 274:1332.

\bibitem[Maksym and Chakraborty, 1990]{MaksymPRL}
Maksym, P.~A. and Chakraborty, T. (1990).
\newblock {\em Phys.\ Rev.\ Lett.}, 64:108.

\bibitem[Maksym et~al., 2000]{MaksymReview}
Maksym, P.~A., Imamura, H., Mallon, G.~P., and Aoki, H. (2000).
\newblock {\em J.\ Phys.: Condens.~Matter}, 12:R299.

\bibitem[Merkt et~al., 1991]{Merkt91}
Merkt, U., Huser, J., and Wagner, M. (1991).
\newblock Energy spectra of two electrons in a harmonic quantum dot.
\newblock {\em Phys.\ Rev.\ B}, 43:7320.

\bibitem[Oh et~al., 1996]{Oh96}
Oh, J.~H., Chang, K.~J., Ihm, G., and Lee, S.~J. (1996).
\newblock Electronic structure and optical properties of coupled quantum dots.
\newblock {\em Phys.\ Rev.\ B}, 53:R13264.

\bibitem[Partoens et~al., 1997]{Partoens97}
Partoens, B., Schweigert, V.~A., and Peeters, F.~M. (1997).
\newblock {\em Phys.\ Rev.\ Lett.}, 79:3990.

\bibitem[Pfannkuche et~al., 1993]{Pfannkuche93}
Pfannkuche, D., Gudmundsson, V., and Maksym, P.~A. (1993).
\newblock Comparison of a hartree, a hartree-fock, and an exact treatment of
  quantum-dot helium.
\newblock {\em Phys.\ Rev.\ B}, 47:2244--2250.

\bibitem[Pi et~al., 2001]{Pi01}
Pi, M., Emperador, A., Barranco, M., Garcias, F., Muraki, K., Tarucha, S., and
  Austing, D.~G. (2001).
\newblock Dissociation of vertical semiconductor diatomic artificial molecules.
\newblock {\em Phys.\ Rev.\ Lett.}, 87:066801.

\bibitem[Rontani et~al., 2002]{Rontani02}
Rontani, M., Goldoni, G., Manghi, F., and Molinari, E. (2002).
\newblock Raman signatures of classical and quantum phases in artificial
  molecules.
\newblock {\em Europhys.\ Lett.}, 58:555.

\bibitem[Rontani et~al., 2003]{RontaniNato}
Rontani, M., Goldoni, G., and Molinari, E. (2003).
\newblock {\em New directions in mesoscopic physics (towards nanoscience)}.
\newblock NATO ASI series B: physics. Kluwer, Dordrecht.
\newblock see also cond-mat/0212626.

\bibitem[Rontani et~al., 2001]{Rontani01}
Rontani, M., Troiani, F., Hohenester, U., and Molinari, E. (2001).
\newblock Quantum phases in artificial molecules.
\newblock {\em Solid State Commun.}, 119:309--321.
\newblock Special Issue on Spin Effects in Mesoscopic Systems.

\bibitem[Szafran et~al., 2003]{Szafran03prb}
Szafran, B., Bednarek, S., and Adamowski, J. (2003).
\newblock Correlation effects in vertical gated quantum dots.
\newblock {\em Phys.\ Rev.\ B}, 67(11):115323.

\bibitem[Tarucha et~al., 1996]{Tarucha96}
Tarucha, S., Austing, D.~G., Honda, T., van~der Hage, R.~J., and Kouwenhoven,
  L.~P. (1996).
\newblock {\em Phys.\ Rev.\ Lett.}, 77:3613.

\bibitem[Troiani et~al., 2000]{FilippoQIP}
Troiani, F., Hohenester, U., and Molinari, E. (2000).
\newblock Exploiting exciton-exciton interactions in semiconductor quantum dots
  for quantum-information processing.
\newblock {\em Phys.\ Rev.\ B}, 62:RC2263.

\bibitem[Troiani et~al., 2002]{FilippoBiexciton}
Troiani, F., Hohenester, U., and Molinari, E. (2002).
\newblock Electron-hole localization in coupled quantum dots.
\newblock {\em Phys.\ Rev.\ B}, 65:RC161301.

\end{thebibliography}

\end{document}